\documentstyle[epsfig,onecolumn]{mn}
\begin{document}

\title{Phase Correlations in Cosmic Microwave Background Temperature Maps}

\author[P. Coles, P. Dineen, J. Earl \& Dean Wright] {Peter
Coles, Patrick Dineen, John Earl and Dean Wright\\ School of
Physics \& Astronomy, University of Nottingham, University Park,
Nottingham NG7 2RD, UK.}

%\date{Accepted 1996 ???? ???; Received 1996 ???? ???;
%in original form 1996 ???? ??}
\maketitle

\begin{abstract}
We study the statistical properties of spherical harmonic modes of
temperature maps of the cosmic microwave background. Unlike other
studies, which focus mainly on properties of the amplitudes of
these modes, we look instead at their phases. In particular, we
present a simple measure of phase correlation that can be
diagnostic of departures from the standard assumption that
primordial density fluctuations constitute a statistically
homogeneous and isotropic Gaussian random field, which should
possess phases that are uniformly random on the unit circle. The
method we discuss checks for the uniformity of the distribution of
phase angles using a non-parametric descriptor based on the use
order statistics, which is known as Kuiper's statistic. The
particular advantage of the method we present is that, when
coupled to the judicious use of Monte Carlo simulations, it can
deliver very interesting results from small data samples. In
particular, it is useful for studying the properties of spherical
harmonics  at low $l$ for which there are only small number of
independent values of $m$ and which therefore furnish only a small
number of phases for analysis. We apply the method to the COBE-DMR
and WMAP sky maps, and find departures from uniformity in both. In
the case of WMAP, our results probably reflect Galactic
contamination or the known variation of signal-to-noise across the
sky rather than primordial non-Gaussianity.
\end{abstract}


\section{Introduction}

A crucial ingredient of many cosmological models is the idea that
galaxies and large-scale structure in the Universe grew by a
process of gravitational instability from small initial
perturbations. In the most successful versions of this basic idea,
the primordial fluctuations that seeded this process  were
generated during a period of inflation which, in its simplest
form, is expected to produce fluctuations with relatively simple
statistical properties (Starobinsky 1979, 1980, 1982; Guth 1980;
Guth \& Pi 1981; Linde 1982; Albrecht \& Steinhardt 1982). In
particular the primordial density field in these models is taken
to form a statistically homogeneous (i.e. stationary) Gaussian
random field (Bardeen et al. 1986). This basic paradigm for
structure formation has survived numerous observational
challenges, and has emerged even stronger after recent
confrontations with the 2dF Galaxy Redshift Survey (2dFGRS;
Percival et al. 2001) and the Wilkinson Microwave Anisotropy Probe
(WMAP; Hinshaw et al. 2003). So successful has the standard
paradigm now become that many regard the future of cosmology as
being largely concerned with improving estimates of the parameters
of this basic model rather than searching for alternatives.

There are, however, a number of suggestions that this confidence
in the model may be misplaced and the focus on parameter
estimation may be somewhat premature. For example, the WMAP data
have a number of unusual properties that are not yet completely
understood (Efstathiou 2003; Chiang et al. 2003; Dineen \& Coles
2003; Eriksen et al. 2003). Among the possibilities suggested by
these anomalies is that the cosmic microwave background (CMB)  sky
is not statistically homogeneous and isotropic, and perhaps not
Gaussian either. The latter possibility would be particularly
interesting as it might provide indications of departures from the
simplest versions of inflation (e.g. Linde \& Mukhanov 1997;
Contaldi, Bean \& Magueijo 2000; Martin, Riazuelo \& Sakellariadou
2000; Gangui, Pogosian \& Winitzki 2001; Gupta et al. 2002;
Gangui, Martin \& Sakellariadou 2002; Bartolo, Matarrese \& Riotto
2002). Whether WMAP is hinting at something primordial or whether
there are systematic problems with the data or its interpretation is
unclear. Either way, these suggestions, as well as general
considerations of the nature of scientific method, suggest that it
is a good time to stand back from the prevailing paradigm and look
for new methods of analysis that are capable of testing the core
assumptions behind it rather than taking them for granted.

In this paper we introduce a new method for the statistical
analysis of all-sky CMB maps which is complementary to usual
approaches based on the power--spectrum but which also furnishes a
simple and direct test of the statistical assumptions underpinning
the standard cosmological models. Our method is based on the
properties of the phases of the (complex) coefficients obtained
from a spherical harmonic expansion of all--sky maps. The
advantages of this approach are that it hits at the heart of the
``random--phase'' assumption essential to the definition of
statistically homogeneous and isotropic Gaussian random fields.
Perhaps most importantly from a methodological point of view, is
intrinsically non--parametric and consequently makes minimal
assumptions about the data.

The layout of this paper is as follows. In the next Section we
discuss some technical issues relating to the properties of
spherical harmonic phases which are necessary for an understanding
of the analysis we present. In Section 3 we explain a practical
procedure for assessing the presence of a particular form of
departure from the random phase hypothesis using Kuiper's
statistic. In Section 4 we discuss results obtained by applying
the method to COBE-DMR maps and data from WMAP, as well as some
toy examples. We discuss the outcomes and outline ideas for future
work in Section 5.

\section{Technical Preamble}

\subsection{Fourier Phases}

The method we discuss in this paper is based on recent results
arising from the study of correlations between Fourier domain for
random fields $\delta({\bf x})$ defined over flat two-- or
three--dimensional spaces. In two dimensions, which is the case
closest to the spherical expansion we use in this paper, the
Fourier expansion is of the form
\begin{equation}
\delta({\bf x})=\delta(x,y)=\sum_{k_x}\sum_{k_y} \tilde{\delta}({\bf
k})\exp(i{\bf k}\cdot {\bf x}),
\end{equation}
with ${\bf k}=(k_x,k_y)$ and the Fourier transform $\delta({\bf
k})$ is complex, i.e. \begin{equation} \tilde{\delta}({\bf
k})=|\tilde{\delta}({\bf k})| \exp [i\phi(k_x,k_y)].
\end{equation}
The quantity $\phi({\bf k})$ is the phase of the mode
corresponding to wavevector ${\bf k}$. A two--dimensional flat surface
is a reasonable approximation to a small patch of the sky, so a
Fourier approach has been taken in studies of high-resolution CMB
maps (Bond \& Efstathiou 1987; Coles \& Barrow 1987). The
extension to three-dimensions is trivial.

In orthodox cosmologies, initial density fluctuations constitute a
statistically homogeneous and isotropic Gaussian random field
(Bardeen et al. 1986). Strictly speaking this means that the real
and imaginary parts of $\delta({\bf k})$ are drawn independently
from Gaussian distributions with a variance that depends only on
$|{\bf k}|$. If this assumption holds, the phases $\phi({\bf k})$
are independent and uniformly random on the interval $[0,2\pi]$;
for more details see Watts \& Coles (2003). On the other hand, the
central limit theorem comes into play whenever a large number of
independent Fourier modes are involved in the superposition
represented by Equation (1). As a result, the random--phase
hypothesis on its own suffices to define a weak form of
Gaussianity (Bardeen et al. 1986). As long as the evolution of
density perturbations is linear, initially random phases will
remain so. But in the non-linear regime, mode--mode interactions
can introduce non--randomness into the distribution of phases.
Characterizing non-linear clustering growth using phase
information has been a challenge for a number of years (Peebles
1980; Ryden \& Gramman 1991; Scherrer, Melott \& Shandarin 1991;
Soda \& Suto 1992; Jain \& Bertschinger 1996, 1998), but recently
there have been some important advances in developing practical
methods based on the use of Fourier phases (Chiang \& Coles 2000;
Coles \& Chiang 2000; Chiang 2001; Chiang, Coles \& Naselsky 2002;
Watts, Coles \& Melott 2003; Matsubara 2003; Hikage, Matsubara \&
Suto 2003) and also on relating phase information to more
traditional statistical approaches based on quantities such as the
bispectrum (Watts \& Coles 2003).

One of the difficulties encountered when constructing methods
based on phase information is how to construct quantitative
measures of association appropriate for a circular measure such as
$\phi({\bf k})$. Chiang \& Coles (2000) pointed out that
traditional measures of covariance of the form $\langle XY
\rangle$ between variates $X$ and $Y$ do not work for two phases
$\phi({\bf k}_1)$ and $\phi({\bf k}_2)$ because of the circular
nature of phase angles. They also considered the obvious
generalization to expectations of the form
\begin{equation}
\langle \exp[i\phi({\bf k}_1)] \exp [ i\phi({\bf k}_2)] \rangle =
\langle \exp[i(\phi_1 + \phi_2)] \rangle, \end{equation} for any
two wavevectors ${\bf k_1}$ and ${\bf k_2}$, except for the case
${\bf k_1}=-{\bf k_2}$. Such correlations are zero for any random
field that is statistically homogeneous and isotropic, i.e. one
that has statistical properties that are both translation and
rotation invariant, whether or not it is Gaussian. This
essentially means that the sum of phases $\phi_1+\phi_2$ is also
random; a similar argument can be made for phase differences
$\phi_1-\phi_2$. The case ${\bf k_1}=-{\bf k_2}$ is excluded from
the above argument because the requirement that the original field
$\delta$ is {\em real} imposes the constraint that
\begin{equation} \delta ({\bf k})=\delta(-{\bf k})^*
\end{equation}
so that $\phi({\bf k})=-\phi(-{\bf k})$, $\phi_1+\phi_2$ is
identically zero and the expectation value in Equation (3) is
unity.

One of the useful tricks developed by Chiang \& Coles (2000) for
the analysis of numerical $N$--body simulations was to consider
phase differences instead of the phases themselves. Such
simulations typically employ periodic boundaries and a finite
grid, so the sum in equation (5) is taken over equally-spaced
discrete wavenumbers up to some maximum value determined by the
size of the periodic box. Taking the spacing to be unity one can
most simply construct phase differences along one direction in
${\bf k}$-space, such as
\begin{equation}
D_k = \phi(k+1)-\phi(k)=\arg[\delta(k+1)\delta^*(k)];
\end{equation}
a straightforward generalization of this approach leads to a
technique known as {\em phase mapping} (Chiang, Coles \& Naselsky
2002; Chiang, Naselsky \& Coles 2002) but we shall stick with the
simple form given in equation (5). On top of its simplicity this
approach has a number of advantages. One is that although
translating the coordinate system by an amount $x$ shifts the
phases by $kx$, the phase difference changes by the same amount
for each wavevector. Secondly, if the phases are uniformly random
on the interval $[0,2\pi]$, then the phase differences are also
uniformly random on the same interval. Finally, the behaviour of
the phase differences is relatively simple to visualize, and
interpret in terms of the dynamics of non-linearly evolving
structure (Coles \& Chiang 2000).

However the preceding argument seems to demonstrate that
statistical homogeneity and isotropy require that $<\exp iD_k >=0$
or, in other words, that the phase differences are random. So how
can $D_k$ be  useful statistically? The answer to this question is
that although the expectation value taken over the ensemble of
possible configurations of a random field may indeed be zero, any
average taken over a finite region within single realization will
not correspond to this value. Departures from the global
expectation value contain information about higher--order
fluctuations. In the case of non--linear gravitational clustering
developing within a finite numerical box, what is key is the scale
of the non-linearity compared with the size of the box. If
non-linearity exists only on small scales, then the $D_k$ are very
close to uniform on the unit circle and the box average is close
to the ensemble average (``the box is a fair sample''). If the
scale of structure is close to the scale of the box then $D_k$
becomes extremely non-uniform (Coles \& Chiang 2000). Using a
bigger box actually makes it harder to characterize non-linear
structure on a given scale using phase differences. For related
comments see Hikage, Matsubara \& Suto (2003).

\subsection{Spherical Harmonic Phases}

We can describe the distribution of fluctuations in the microwave
background over the celestial sphere using a sum over a set of
spherical harmonics:
\begin{equation}
\label{deltatovert} \Delta (\theta, \phi)= \frac{(T(\theta ,\phi
)-\bar{T})}{\bar{T}}=\sum _{l=1}^{\infty }\sum _
{m=-l}^{m=+l}a_{l,m}Y_{lm}(\theta ,\phi ).
\end{equation}
Here $\Delta(\theta,\phi)$ is the departure of the temperature
from the average at angular position $(\theta ,\phi)$ on the
celestial sphere in some coordinate system, usually Galactic. The
$Y_{lm}(\theta ,\phi )$ are spherical harmonic functions which we
define in terms of the Legendre polynomials $P_{lm}$ using
\begin{equation}
Y_{lm}(\theta ,\phi )= (-1)^m
\sqrt{\frac{(2l+1)(l-m)!}{4\pi(l+m)!}} P_{lm}(\cos \theta)
\exp(im\phi),
\end{equation}
i.e. we use the Condon-Shortley phase convention. In Equation (6),
the $a_{l,m}$ are complex coefficients which can be written
\begin{equation}
a_{l,m}=|a_{l,m}|\exp[i\phi_{l,m}].
\end{equation}
Note that, since $\Delta$ is real, the definitions (7) and (8)
require the   following relations between the real and imaginary
parts of the $a_{l,m}$: if $m$ is odd then
\begin{equation}
\Re (a_{l,m})  =  - \Re(a_{l,-m}), \,\,\,\,\, \Im(a_{l,m})  =
\Im(a_{l,-m});
\end{equation}
while if $m$ is even
\begin{equation}
\Re (a_{l,m})  = \Re(a_{l,-m}), \,\,\,\,\, \Im(a_{l,m})  =  -
\Im(a_{l,-m});
\end{equation}
and if $m$ is zero then
\begin{equation}
\Im(a_{l,m}) = 0.
\end{equation}
From this it is clear that the $m=0$ mode always has zero phase,
and there are consequently only $l$ independent phase angles
describing the harmonic modes at a given $l$.

If the primordial density fluctuations form a Gaussian random
field as described in Sec. 2.1, then the temperature variations
induced across the sky form a Gaussian random field over the
celestial sphere. By analogy with the above discussion above, this
means that
\begin{equation}
\langle a_{l,m}a_{l',m'}^* \rangle = C_l \delta_{ll'}\delta_{mm'},
\end{equation}
where $C_l$ is the angular power spectrum, the subject of much
scrutiny in the context of the cosmic microwave background (e.g.
Hinshaw et al. 2003), and $\delta_{xx'}$ is the Kronecker delta
function. Since the phases are random, the stochastic properties
of a statistically homogeneous and isotropic Gaussian random field
are fully specified by the $C_l$, which determines the variance of
the real and imaginary parts of $a_{l,m}$ both of which are
Gaussian.

To look for signs of phase correlations in CMB maps one can begin
with a straightforward generalization of the ideas introduced
above for Fourier phases. There are, however, some technical
points to be mentioned. First, the set of spherical harmonics, and
consequently the phases of the harmonic coefficients, is defined
with respect to a particular coordinate system. This is also true
in the Fourier domain but, as we discussed above, dealing with
translations of the coordinate system in that case is relatively
straightforward because $k_x$ and $k_y$ are equivalent. This is
not so for spherical harmonics: the definitions of $l$ and $m$ are
quite different, and the distinction is introduced when the
$z$-axis of a three-dimensional coordinate system is chosen to be
the polar axis. Changing this axis muddles up the spherical
harmonic coefficients in a much more complex way than is the case
for Fourier modes.

Another point worth stressing is that the sphere is a finite
space, so when analyzing the statistical properties of maps on a
single sky one never has an exact ensemble average. This is less
of an issue when dealing with spherical harmonic modes at high
$l$, because the average over a single sky is then close to that
over the probability distribution, but at low $l$ there is the
perennial problem of so-called ``cosmic variance''. Suppose one
were to construct phase differences in the manner of Section 2.1,
i.e.
\begin{equation}
D_m(l)=\phi_{l,m+1}-\phi_{l,m}. \end{equation} Then the
distribution of the $D_m$ for high $l$ should tend to uniform if
the field is statistically homogeneous over the sphere whether or
not the temperature fluctuations are Gaussian. However, at low
$l$, the finiteness of the sky come into play and the distribution
over the small number of modes available will depart from the
uniform case. These fluctuations at low $l$ depend on
higher--order statistical information about the form of the
fluctuations, so in this case cosmic variance can actually be
helpful.

We shall explain how we cope with this aspect of phases and the
behaviour in more detail in Section 3. For the time being we note
that, if the orthodox cosmological interpretation of temperature
fluctuations is correct, the phases of the $a_{l,m}$ should be
random and so should phase differences of the form
$\phi_{l,m+1}-\phi_{l,m}$ and $\phi_{l+1,m}-\phi_{l,m}$. The aim
of this paper is introduce a method of checking whether this is
so.

\subsection{Departures from Standard Statistics}

In studying the properties of phase correlations of CMB maps, our
intention is to characterize departures from the standard
cosmological framework. The usual hypothesis, that primordial
density fluctuations form a statistically homogeneous and
isotropic Gaussian random field, results in uncorrelated phases.
Most discussions of phase correlations in the literature refer to
them as diagnostic of non--Gaussianity in some form. What this
actually means is usually that the field in question is not a
homogeneous and isotropic Gaussian random field. Fields can be
defined which are non--Gaussian but are both homogeneous and
isotropic; Coles \& Barrow (1987) give examples. Such fields
certainly do not have random phases, but the form of phase
association that characterizes them can be quite complex (Watts \&
Coles 2003; Matsubara 2003; Hikage, Matsubara \& Suto 2003).
However, Watts \& Coles (2003) established an important connection
between the form of non--Gaussianity exhibited by such fields and
the bispectrum. The bispectrum is now a standard part of the
cosmologists armoury for studies of large--scale structure in
three dimensions (e.g. Peebles 1980; Luo 1994; Heavens 1998;
Matarrese, Verde \& Heavens 1997; Scoccimarro et al. 1998;
Scoccimarro et al. 1998; Scoccimmarro, Couchman \& Frieman 1999;
Verde et al. 2000, 2001, 2002) and for CMB studies (Ferreira,
Magueijo \& Gorski 1998; Heavens 1998; Magueijo 2000; Contaldi et
al. 2000; Phillips \& Kogut 2001; Sandvik \& Magueijo 2001; Santos
et al. 2001; Komatsu et al. 2002; Komatsu et al. 2003). The
bispectrum, being essentially the three--point correlation
function in harmonic space, can be generalized to higher order
polyspectra (Stirling \& Peacock 1996; Verde \& Heavens 2001; Hu
2001; Cooray 2001). However the {\em estimation} of these
polyspectra generally involves taking averages that assume that
the field one is dealing with is indeed statistically homogeneous
and  they are therefore not in themselves tests of that
hypothesis.

On the other hand, it is also possible for fields to be Gaussian
in some sense, but not strictly Gaussian in the sense we defined
in Sec. 2.1. As a consequence of strict Gaussianity, all the
$n$-dimensional joint probability distributions of the field
values at $n$ different spatial locations are multivariate
Gaussian (Bardeen et al. 1986). Fields can be constructed in
which, say, the one--point distribution is Gaussian but the
higher--order joint distributions are not. These would be Gaussian
in some sense, but again would not have random phases and may or
may not be statistically homogeneous and isotropic. In the Fourier
or spherical harmonic domain, one can construct fields in which
the harmonic coefficients are Gaussian but not independent. Such
fields are Gaussian, but either statistically inhomogeneous or
statistically anisotropic and would have non-random phases.

Any or all of these deviations from simple Gaussianity might be
expected to occur in CMB in various circumstances. Statistical
inhomogeneity and/or anisotropy over the sky might result from
incorrectly subtracted foreground emission from the Galaxy.
Alternatively, the scanning pattern used to produce all-sky map
will generally not sample the sky uniformly resulting in a
signal--to--noise that varies over the celestial sphere. Perhaps
more exotically, compact universe models with a non-trivial
topology should result in repeated features in the temperature
pattern on the sky (Levin, Scannapieco \& Silk 1998; Scannapieco,
Levin \& Silk 1999; Levin 2002; Rocha et al. 2002). All these
possibilities should result in some form of phase correlation, and
indeed there is already some evidence from the preliminary release
of the WMAP data (Chiang et al. 2003; Eriksen et al. 2003).

The focus of most discussions of higher-order CMB statistics is
the possible detection of primordial non--Gaussianity. As we
discussed in the introduction, most versions of the inflationary
universe idea produce primordial fluctuations that are extremely
close to the strictly Gaussian form. However, as we mentioned in
the Introduction, there are circumstances in which the primordial
fluctuations could be non--Gaussian. In such cases there will be
information concerning the level of non--Gaussianity in the
distribution of phases, but this usually appears in a form that is
picked up by the bispectrum (Watts \& Coles 2003).

Clearly there are many different possible forms of departure from
the standard statistical assumption and consequently many possible
signatures in the distribution of phases. Our aim in this paper is
to look for a simple method that is a powerful complement to the
usual approach via the polyspectra and hopefully sheds some light
on previous debates about departures from Gaussianity in the
COBE--DMR data (Ferreira, Magueijo \& Gorski 1998; Pando,
Valls-Gabaud \& Fang 1998; Bromley \& Tegmark 1999; Magueijo
2000). As we shall see, a test based on the distribution of
harmonic phases and/or phase differences fits the bill rather
nicely.

\section{Testing for Phase Correlations}

\subsection{Kuiper's Statistic}

The approach we take in this paper is to assume that we have
available a set of phases $\phi_{l,m}$ corresponding to a set of
spherical harmonic coefficients $a_{l,m}$ obtained from a data
set, either real or simulated. We can also form from these phases a set of phase differences as described in the previous section.
Let us assume, therefore, that we have $n$ generic angles,
$\theta_1 \ldots \theta_n$. Under the standard statistical
assumption these should be random, apart from the constraints
described in Section 2.2. The first thing we need is a way of
testing whether a given set of phase angles is consistent with
being drawn from uniform distribution on the unit circle. This is
not quite as simple as it seems, particularly if one does not want
to assume any particular form for actual distribution of angles,
such as a bias in a particular direction; see Fisher (1993).
Fortunately, however, there is a fully non--parametric method
available, based on the theory of order statistics, and known as
as Kuiper's statistic (Kuiper 1960).

Kuiper's method revolves around the construction of a statistic,
$V$, obtained from the data via the following prescription. First
the angles are sorted into ascending order, to give the set
$\{\theta _{1},\ldots ,\theta _{n}\}$. It does not matter whether
the angles are defined to lie in $[0,2\pi]$, $[-\pi,+\pi]$ or
whatever. Each angle $\theta _{i}$ is divided by $2\pi$ to give a
set of variables $X_{i}$, where $i=1\ldots n$. From the set of
$X_i$ we derive two values $S^+_{n}$ and $S^-_{n}$ where
\begin{equation}
S^{+}_{n} = {\rm max}
\left\{\frac{1}{n}-X_{1},\frac{2}{n}-X_{2},\ldots ,1-X_{n}\right\}
\end{equation}
and   \begin{equation} S^{-}_{n} = {\rm max}
\left\{X_{1},X_{2}-\frac{1}{n},\ldots
\dot{,}X_{n}-\frac{n-1}{n}\right\}.
\end{equation}
Kuiper's statistic, $V$, is then defined as
\begin{equation}
\label{TestStatisticV} V=(S^{+}_{n}+S^{-}_{n})\cdot
\left(\sqrt{n}+0.155+\frac{0.24}{\sqrt{n}}\right).
\end{equation}
Anomalously large values of $V$ indicate a distribution that is
more clumped than a uniformly random distribution, while low
values mean that angles are more regular. The test statistic is
normalized by the number of variates, $n$, in such a way that
standard tables can be constructed to determine significance
levels for any departure from uniformity; see Fisher (1993). In
this context, however, it is more convenient to determine
significance levels using Monte Carlo simulations of the ``null''
hypothesis of random phases. This is partly because of the large
number of samples available for test, but also because we can use
them to make the test more general.

The first point to mention is that a given set of phases, say
belonging to the modes at fixed $l$ is not strictly speaking
random anyway, because of the constraints noted in equations
(9)--(11). One could deal with this by discarding the conjugate
phases, thus reducing the number of data points, but there is no
need to do this when one can instead build the required symmetries
into the Monte Carlo generator.

In addition, suppose the phases of the temperature field over the
celestial sphere were indeed random, but observations were
available only over apart of the sky, such as when a Galactic cut
is applied to remove parts of the map contaminated by foregrounds.
In this case the mask may introduce phase correlations into the
observations so the correct null hypothesis would be more
complicated than simple uniform randomness. As long as any such
selection effect were known, it could be built into the Monte
Carlo simulation. One would then need to determine whether $V$
from an observed sky is consistent with having been drawn from the
set of values of $V$ generated over the Monte Carlo ensemble.

There is also a more fundamental problem in applying this test to
spherical harmonic phases. This is that a given set of $a_{l,m}$
depends on the choice of a particular coordinate axis.  A given
sky could actually generate an infinite number of different sets
of $\phi_{l,m}$ because the phase angles are not rotationally
invariant. One has to be sure to take different choices of
$z$-axis into consideration when assessing significance levels, as
a random phase distribution has no preferred axis while systematic
artifacts may. A positive detection of non--randomness may result
from a chance alignment of features with a particular coordinate
axis in the real sky unless this is factored into the Monte Carlo
simulations to. For both the real sky and the Monte Carlo skies we
therefore need not a single value of $V$ but a distribution of
$V$-values obtained by rotating the sky over all possible angles.
A similar approach is taken by Hansen, Marinucci \& Vittorio
(2003). This method may seem somewhat clumsy, but a test is to be
sensitive to departures from statistical homogeneity one should
not base the test on measures that are rotationally invariant,
such as those suggested by Ferreira, Mageuijo \& Gorski (1998) as
these involve averaging over the very fluctuations one is trying
to detect.

\subsection{Rotating  the \protect\( a_{l,m}\protect \)}

In view of the preceding discussion we need to know how to
transform a given set of $a_{l,m}$ into a new set when the
coordinate system is rotated into a different orientation. The
method is fairly standard, but we outline it here to facilitate
implementation of our approach.

Any rotation of the cartesian coordinate system \(
S\{x,y,z\}\mapsto S^{\prime }\{x,y,z\} \) can be described using a
set of three Euler angles $\alpha$, $\beta$, $\gamma$, which
define the magnitude of successive rotations about the coordinate
axes. In terms of a rotation operator $\hat{D}(\alpha, \beta,
\gamma)$, defined so that a field $f(r,\theta,\phi)$ transforms
according to
\begin{equation}
\hat{D}(\alpha, \beta, \gamma) f(r,\theta,\phi) =
f'(r,\theta,\phi)=f(r,\theta', \phi'),
\end{equation}
a vector ${\bf r}$ is transformed as
\begin{equation}
{\bf r}'={\bf D}(0,0,\gamma) {\bf D} (0,\beta,0) {\bf D}(\alpha,
0, 0) {\bf r} \equiv{\bf D}(\alpha, \beta, \gamma) {\bf r}.
\end{equation}
Here ${\bf D}$ is a matrix representing the operator $\hat{D}$,
i.e.
\begin{equation}
{\bf D}(\alpha, \beta, \gamma) = \left( \begin{array}{ccc} \cos
\gamma & \sin \gamma & 0 \\ -\sin \gamma  & \cos \gamma & 0\\ 0 &
0 & 1 \end{array} \right)\left( \begin{array}{ccc} \cos \beta & 0
& -\sin\beta \\ 0  & 1 & 0\\ \sin \beta & 0 & \cos \beta
\end{array} \right)\left( \begin{array}{ccc} \cos
\alpha & \sin \alpha & 0 \\ -\sin \alpha  & \cos \alpha & 0\\ 0 &
0 & 1
\end{array} \right).
\end{equation}
The Wigner \( D \) functions describe the rotation operator used
to realise the transformations of covariant components of tensors
with arbitrary rank of rank \( l \). The functions, written as \(
D^{l}_{m,m^{\prime }} \), transform a tensor from \( S\{x,y,z\} \)
to \( S^{\prime }\{x^{\prime },y^{\prime },z^{\prime }\} \).
Consider a tensor \( Y_{l,m}(\theta ,\phi ) \) defined under the
coordinate system \( S \) and apply the rotation operator we get:
\begin{equation}
\label{Wigner1} \hat{D}(\alpha ,\beta ,\gamma )Y_{l,m}(\theta
,\phi )=Y_{l,m^{\prime }}(\theta ^{\prime },\phi ^{\prime })=\sum
_{m}Y_{l,m}(\theta ,\phi )D^{l}_{m,m^{\prime }}(\alpha ,\beta, \gamma )
\end{equation}
This means that the transformation of the tensor under the
rotation of the coordinate system can be represented as a matrix
multiplication. An alternative representation of the \( D \)
functions \( D^{l}_{m,m^{\prime }} \) in terms of the Euler angles
is
\begin{equation}
\label{wignerlittled} D^{l}_{m,m^{\prime }}(\alpha ,\beta ,\gamma
)=e^{-im\alpha }d^{l}_{m,m^{\prime }}(\beta )e^{-im^{\prime
}\gamma }.
\end{equation}
In this representation the parts of the \( D \) function depending
upon the individual Euler angles are separated. The \(
d^{l}_{m,m^{\prime }} \) form the elements of a rotation matrix
and only depend upon the angle \( \beta  \). Using this
representation makes calculating the values of the \(
D^{l}_{m,m^{\prime }} \) easier since the \( d^{l}_{m,m^{\prime }}
\) possess a number of symmetry properties meaning that although
one quarter of the elements need to be directly calculated the
values of the remainder can be inferred.
\begin{figure} {\centering
\epsfig{file=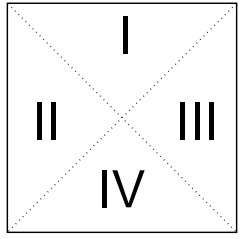} \par}
\caption{\label{fig: symmetries}Mapping the symmetries (see
text).}
\end{figure}
Figure (\ref{fig: symmetries}) shows how the following symmetries
can be used to map the elements in the top quadrant of the
rotation matrix into each of the other quadrants. First,
\begin{equation}
d^{l}_{m,m^{\prime }}=(-1)^{m-m^{\prime }}d^{l}_{m,m^{\prime }},
\end{equation}
which maps elements in region I into region II as shown in Figure
1. Next, \begin{equation} d^{l}_{m,m^{\prime }}=d^{l}_{-m^{\prime
},-m}, \end{equation} which maps elements in region I into region
III. And finally, \begin{equation}
d^{l}_{m,m^{\prime}}=(-1)^{m-m^{\prime }}d^{l}_{-m,-m^{\prime }}.
\end{equation} This maps elements in region I into region IV.
The relations (22) to (24) can be used in conjunction with
Equation (\ref{wignerlittled}), to find all the elements of the
Wigner \( D \) functions from the values of \( d^{l}_{m,m^{\prime
}} \) in the upper quadrant. Similar, but more complicated,
symmetry relations exist for the \( D^{l}_{m,m^{\prime }} \).

For a given value of  \(l\) we can construct a sum of spherical
harmonics in the coordinate system \( S^{\prime }\{r,\theta
^{\prime },\phi ^{\prime }\} \) as follows
\begin{equation}
\label{Wigner2} Y_{l}^{\prime }=\sum _{m^{\prime }}a^{\prime
}_{l,m^{\prime }} Y_{l,m^{\prime }}(\theta ^{\prime },\phi
^{\prime }).
\end{equation}
We can use this and the transformation in equation (\ref{Wigner1})
to find the transformed sum, \( Y_{l} \) in the coordinate system
\( S\{r,\theta ,\phi \} \)
\begin{eqnarray}
Y_{l} & = & \sum _{m^{\prime }}a^{\prime }_{l,m^{\prime }}\left(
\sum _{m}D^{l}_{m,m^{\prime }}(\alpha ,\beta ,\gamma
)Y_{l,m}(\theta ,\phi )\right) \nonumber \\
 & = & \sum _{m}\left( \sum _{m^{\prime }}a_{l,m^{\prime }}
 D^{l}_{m,m^{\prime }}(\alpha ,\beta ,\gamma )\right) Y_{l,m}(\theta ,\phi )=
 \sum _{m}a_{l,m}Y_{l,m}(\theta ,\phi )\label{Wigner3}
\end{eqnarray}
This gives an expression for the rotated coefficients
\begin{equation}
\label{RotatedCoefficients} a_{l,m}=\sum _{m^{\prime
}}a_{l,m^{\prime }}D^{l}_{m,m^{\prime }}(\alpha ,\beta ,\gamma ).
\end{equation}
Finding the rotated coefficients therefore requires a simple
matrix multiplication once the appropriate \( D \) function is
known. To apply this in practice one needs a fast and accurate way
of generating the matrix elements \( D^{l}_{m,m^{\prime }} \) for
the rotation matrix. There are \( (2l+1)^{2} \) elements needed to
describe the rotation of each mode and the value of each element
depends upon the particular values of \( (\alpha ,\beta ,\gamma )
\)used for that rotation.

Varshalonich, Moskalev \& Khersonskii (1988) provide a set of
matrices for the \( d^{l}_{m,m^{\prime }}(\beta ) \) for the modes
\( l=1/2,1,3/2,\ldots ,5 \). For example, the case \( l=1 \) is
\begin{equation}
\label{j=1matrix} d^{1}_{m,m^{\prime}}(\beta)
=\left(\begin{array}{cccc}
  \frac{1+\cos \beta }{2} &
-\frac{\sin \beta }{\sqrt{2}} & \frac{1-\cos \beta }{2}\\
\frac{\sin \beta }{\sqrt{2}} & \cos \beta  & -\frac{\sin \beta
}{\sqrt{2}}\\   \frac{1-\cos \beta }{2} & \frac{\sin \beta
}{\sqrt{2}} & \frac{1+\cos \beta }{2}
\end{array}\right).
\end{equation}
Fortunately, rather than working these matrices out by hand,
Varshalovich et al. (1988) includes a set of iteration formulae,
which, in conjunction with the symmetry relations for the \(
d^{l}_{m,m^{\prime }} \), can be used to generate matrices for the
higher modes from the previous ones. The first iteration formula
provides a way of generating an element in the mode \( l+1 \) by
using the equivalent element in the matrices for the modes \( l \)
and \( l-1 \).
\begin{eqnarray}
\label{upiteration} D^{l+1}_{m,m^{\prime }} & = & f\left(
D^{l}_{m,m^{\prime }},D^{l-1}_{m,m^{\prime }}\right) \nonumber\\ &
= & (l+1)(2l+1)\left[ -\frac{\sqrt{(l^{2}-m^{2})(l^{2}-m^{\prime
2})}D^{l-1}_{m,m^{\prime }}+\left( \cos \beta -\frac{m,m^{\prime
}}{l(l+1)}\right) D^{l}_{m,m^{\prime
}}}{\sqrt{((l+1)^{2}-m^{2})((l+1)^{2}-m^{\prime 2})}}\right].
\end{eqnarray}
Applying this formula repeatedly allows the central nine elements
to be generated for all of the matrices.

Another iteration formula uses the values already in the matrix to calculate
the element directly to the left. For example the element \( D^{3}_{02} \)
could be generated using the elements \( D^{3}_{01} \) and \( D^{3}_{00} \).
\begin{eqnarray}
\label{acrossiteration} D^{l}_{m,m^{\prime }+1} & = & f\left(
D^{l}_{m,m^{\prime }},D^{l}_{m,m^{\prime }-1}\right) \nonumber\\ &
= & \frac{2e^{-i\gamma }}{\sqrt{(l-m^{\prime})(l+m^{\prime }+1})}
\times \nonumber\\ & & \times \left[ \frac{(-m+m^{\prime }\cos
\beta )D^{l}_{m,m^{\prime }}}{\sin \beta }-
 \frac{1}{2}\sqrt{(l+m^{\prime })(l-m^{\prime }+1)}D^{l}_{m,m^{\prime }-1}e^{-i\gamma
 }\right].
\end{eqnarray}
Similar iteration functions exist to move up a matrix and across a
matrix to the right.
\begin{figure}
{\centering \epsfig{file=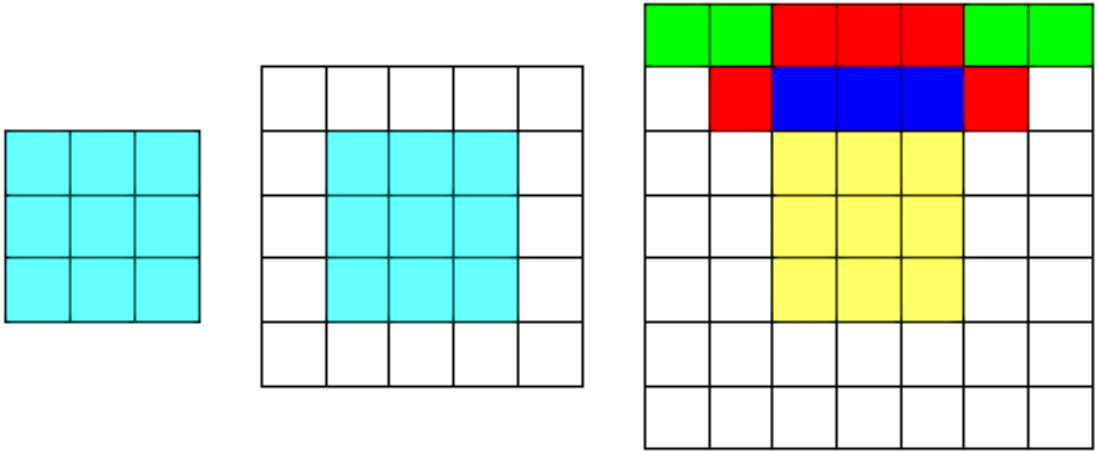, width=8cm,
angle=-90}
\par}\caption{\label{fig: IterationMethod} The basic steps
involved in the iteration method described in the text for
increasing matrix size: (\textit{left}) $l$=1, (\textit{middle})
$l=2$, and (\textit{right}) $l$=3.}
\end{figure}

Figure (\ref{fig: IterationMethod}) shows the iteration scheme.
Iteration formula (\ref{upiteration}) is used to generate the
elements coloured in yellow from the elements coloured in cyan.
The elements coloured in yellow are then used to calculate the
elements coloured in blue. The yellow and blue elements are used
to calculate the elements coloured in red. The red elements are
used to calculate the elements coloured in green. The symmetry
relations are used to calculate the elements in the other three
quadrants. Once the \( l=3 \) matrix has been calculated it can be
used with the \( l=2 \) matrix to find the fourth matrix, and so
on and so forth.

This method is fairly straightforward to implement but care is
needed. The main problem with it is that the second iteration
formula shown, equation (\ref{acrossiteration}) includes a term in
which a factor is divided by \( \sin \beta  \). If \( \beta \simeq
0 \) or \( \beta \simeq \pi  \) then \( \sin \beta \simeq 0 \)
leading to numerical problems.

An alternative approach is to calculate each \( d^{l}_{mm^{\prime
}} \) using factorials. The procedure encoded the definition of
the \( d^{l}_{m,m^{\prime }} \) given in Edmonds (1960):
\begin{eqnarray}
\label{factorialdjmm} d^{l}_{m,m^{\prime }}(\beta ) & = & \left[
\frac{(l+m^{\prime })!(l-m^{\prime })!}{(l+m)!(l-m)!}\right]
^{\frac{1}{2}}\sum _{k}\left( \begin{array}{c} l+m\\ l-m^{\prime
}-k
\end{array}\right) \left( \begin{array}{c}
l-m\\ k
\end{array}\right) (-1)^{l-m^{\prime }-k}\nonumber \\
 &  & \times \left( \cos \beta \right) ^{2k+m^{\prime }+m}
 \left( \sin \beta \right)^{2l-2k-m^{\prime }-m}.
\end{eqnarray}
In this expression the variable \( k \) varies between \( 0 \) and
the smallest of \( l-m \) and \( l-m^{\prime } \). This method
avoids the problem of dividing through by \( (\sin \beta )^{n} \)
when \( \beta  \) is small. The results are far more stable and
satisfy both of the techniques given above for testing the
rotation code. The procedure can be tested using a random number
generator to produce matrices for the full range of possible Euler
angles.

There are two main tests that can be made of the results. First,
the rotation code should generate the correct values for the \(
d^{l}_{m,m^{\prime }} \) matrices given in Varshalovich et al.
(1988). Since the iteration formula are for the \(
D^{l}_{m,m^{\prime }} \) they will be applicable to the \(
d^{l}_{m,m^{\prime }} \) if \( \alpha \) and \( \gamma  \) are set
to zero. Choosing a value of \( \beta \) and using it in matrices
(\ref{Wigner2}) and (\ref{Wigner3}) should reproduce the precise
calculation of the \( l=3,4\textrm{ and }5 \) matrices as given by
Varshalovich et al. (1988). Second, the calculated matrices should
be rotation matrices, so they should preserve the length of a
vector. We define the vector of coefficients \( A_{l} \) for a
particular mode \( l \) as
\begin{equation}
\label{vectoralm} A_{l}=(a_{l,-l},a_{l,-l+1},\ldots
,a_{l,-1},a_{l,0},a_{l,1},\ldots a_{l,l-1},a_{l,l})
\end{equation}
 and use this to define the length of \( A_{l} \) to be
\begin{equation}
\label{lengthvectoralm}
|A_{l}|=\sqrt{\sum _{m}|a_{l,m}|^{2}}.
\end{equation}
 The length should be preserved under the transformation. This allows the rotation
code to be tested for any  values of \( l>5 \) and for \( \alpha
,\gamma \neq 0 \).

\subsection{Implementation}

In order to apply these ideas to make a test of CMB fluctuations,
we first need a temperature map from which we can obtain a
measured set of $a_{l,m}$. Employing equation
\ref{RotatedCoefficients} with some choice of Euler angles yields
a rotated set of the $a_{l,m}$. It is straightforward to choose a
set of angles such that random orientations of the coordinate axis
can be generated. The Wigner $D$ function needed to compute this
transformation is generated using the techniques described in
Section 3.3.  For the purposes of this paper we only compute the
effect of rotation on low values of $l$. There is no difficulty in
principle in extending this method to very high spherical
harmonics, but the computational generation of Wigner $D$ scales
by $\sim l^2$ so have chosen to limit ourselves to $l$=20. Once a
rotated set has been obtained, Kuiper's statistic is calculated
from the relevant transformed set of phases. For each CMB map,
3000 rotated sets are calculated by this kind of resampling of the
original data, producing 3000 values of $V_{\mathrm{cmb}}$. The
values of the statistic are binned to form a measured (re-sampled)
distribution of $V_{\mathrm{cmb}}$. The same procedure is applied
to the 1000 Monte Carlo sets of $a_{l,m}$ drawn from a uniformly
random distribution, i.e. each set is rotated 3000 times and a
distribution of $V_{\mathrm{MC}}$ under the null hypothesis is
produced. These realizations are then binned to created an overall
global average distribution under the null hypothesis. In the
following applications, the distribution of the 3000 values of V
was split into 100 equally-spaced bins ranging from $V=0$ to 3.0
for different $l$ and 100 equally-spaced bins ranging from $V=0$
to 5.2 for different $m$.

In order to determine whether the distribution of
$V_{\mathrm{cmb}}$ is compatible with a distribution drawn from a
sky with random phases, we use a simple $\chi^2$ test, using
\begin{equation}
\chi^2=\sum_{i}\frac{(f_i-\overline{f_i})^2}{\overline{f_i}}
\end{equation}
where the summation is over all the bins  and $\overline{f_i}$ is
the number expected in the $i$th bin from the overall average
distribution. The larger the value of $\chi^2$ the less likely the
distribution functions are to be drawn from the same parent
distribution. Values of $\chi^2_{\mathrm{MC}}$ are calculated for
the 1000 Monte Carlo distributions and $\chi^2_{\mathrm{cmb}}$ is
calculated from the distribution of $V_{\mathrm{cmb}}$. If the
value of $\chi^2_{\mathrm{cmb}}$ is greater than a fraction $p$ of
the values of $\chi^2_{\mathrm{MC}}$, then the phases depart from
a uniform distribution at significance level $p$. We have chosen
95 per cent as an appropriate level for the level at which the
data are said to display signatures that are not characteristic of
a statistically homogeneous Gaussian random field.

\section{Applications}

\subsection{Toy Models}

As a first check on the suitability of our method for detecting
evidence of non-Gaussianity in CMB skymaps, sets of spherical
harmonic coefficients were created with phases drawn from
non-uniform distributions. Three types of distributions were used
to test the method: (i) the phases were coupled (ii) the phases
were drawn from a cardioid distribution and (iii) the phases were
drawn from a wrapped Cauchy distribution; see Fisher (1993). The
fake sets of coefficients were created in a manner that insured
that the orthonormality of the coefficients were preserved as
outlined in equations (9)--(11).
%For the purpose of this test, the
%Monte Carlo realizations were each rotated 3000 times and Kuiper's
%statistic was calculated both for the phases and for the phase
%differences of each rotated set. The distribution of the
%statistics were compared through a \(\chi^2\) test against the
%average distribution for 1000 Monte Carlo skies (whose phases were
%drawn from a uniformly random distribution).

%\begin{figure}
%\includegraphics[height=0.5\textwidth,width=0.5\textwidth,angle=-90]{cardioid.ps}
%\includegraphics[height=0.5\textwidth,width=0.5\textwidth,angle=-90]{wrap.ps}
%\caption{\label{fig:dist} Probability density functions for the
%cardioid and wrapped Cauchy distributions}
%\end{figure}

The first non--uniform distribution of phases is constructed using
the following relationship
\begin{equation}
\label{couple} \phi_{m}=\phi_{m-1}+k
\end{equation}
where \(\phi_{m=-l}\) is a random number chosen between 0 and
2\(\pi\). The results of the simulation with \( k= 1 \) are shown
in panel (i) of Tables \ref{tab:nGpr} and \ref{tab:nGpdr}. The sky
is taken to be non-Gaussian if \(\chi^2_{\rm sky}\) is larger than
95 \% of \(\chi^2_{MC}\) for a particular mode. The phase
difference method should be particularly suited to detecting this
kind of deviation from Gaussianity because the phase differences
are all equal in this case; the distribution of phase differences
is therefore highly concentrated rather than uniform.
Nevertheless, the non-uniformity is so clear that it is evident
from the phases themselves, even for low modes. On the other hand,
assessing non-uniformity from the quadrupole mode $l=2$ on its own
is difficult even in this case because of the small number of
independent variables available.

The cardioid distribution was chosen because studies of phase
evolution in the non-linear regime have shown phases rapidly move
away from their initial values, however, they wrap around many
multiples of 2\(\pi\) and the 'observed' phases appear random
(Chiang and Coles 2000; Watts et al. 2003). This evolution can be
distinguished by the phases differences which appear to be drawn
from a roughly cardioid distribution (Watts et al. 2003). The
cardioid distribution has a probability density function given by
\begin{equation}
\label{card} f(\theta)=\frac{1}{2\pi}[1+2\rho\cos(\theta)],\quad
0\leq\theta<2\pi,0\leq\rho\leq1/2,
\end{equation}
where \(\rho\) is the mean resultant length. As \(\rho
\rightarrow0\) the distribution converges to a uniform
distribution. The kurtosis of the distribution is given by
\begin{equation}
\label{kcard} K=-\frac{\rho^4}{(1-\rho)^2}.
\end{equation}
Sets of coefficients were produced from distributions with
\(\rho=0,0.2,0.4,0.5\). The distributions themselves were
generated using rejection methods. The results are given in panel
(ii) of Tables \ref{tab:nGpr} and \ref{tab:nGpdr}. The results
with \(\rho=0\) are not shown: they indicate the coefficients are
taken from a Gaussian random field, as expected. Furthermore, with
\(\rho=0.5\) the non-uniformity detected in the phases and phase
differences for \(l=20\) is not particularly significant, however,
the kurtosis for this distribution is not especially large.
\begin{table}
\begin{center}
\begin{tabular}{|cc|c|c|c|c|c|c|c|c|c|} \hline\hline && i
&\multicolumn{3}{c|}{ii} &\multicolumn{5}{c|}{iii} \\ \hline & $K$
& &-0.003&-0.07&-0.25&0.06&0.37&1.44&5.76&$\infty$\\ \hline $l$
&&&&&&&&&&\\ 2 && 76  & 79 & 29 & 81 & 53 & 21 & 68
& \textbf{98} & 52
\\ 4 && \textbf{96}& 56 & 41 & 44 & 68 & 77 & 39 & 68 &
\textbf{100} \\ 6 && \textbf{100} & 18 & 56 & 83 & 61 &
41 & \textbf{99} & 85 & \textbf{100} \\ 8 && \textbf{100}
&\textbf{99} & 16 & 64 & \textbf{97} & 18 &
\textbf{99} & \textbf{100} & \textbf{100} \\ 10 && \textbf{100}
& 11 & 64 & 39 & 19 & 63 & 24 & \textbf{99} &
\textbf{100}
\\ 12 && \textbf{100} & 22 & 45 & 34 & 38 & \textbf{96} & 39
& \textbf{95} & \textbf{100} \\ 14 && \textbf{100} & 63 & 4
& 82 & 60 & 62 & 93 &\textbf{97} & \textbf{100} \\ 16 &&
\textbf{100} & 69 & 41 & 31 & 16 & 68 & 72 &
\textbf{100} & \textbf{100} \\ 18 && \textbf{100} & 83 & 63 &
44 & 54 & 18 & \textbf{100} & \textbf{100} & \textbf{100}
\\ 20 && \textbf{100} & 5 & 72 & \textbf{95} & 64 & 1 & 83 &
\textbf{100} & \textbf{100} \\\hline
\end{tabular}
\end{center}
\caption{ \label{tab:nGpr}Results for non-Gaussian skies based on
phases between consecutive values of $m$ at a fixed $l$. The
results show the significance levels of detected departures from
random phases for simulated distributions corresponding to (i)
coupled phases, (ii) a Cardioid distribution and (iii) a wrapped
Cauchy distribution. In the latter two cases various parameter
choices are shown as described in the text. These results are
obtained from 1000 Monte Carlo skies.}
\end{table}

\begin{table}
\begin{center}
\begin{tabular}{|cc|c|c|c|c|c|c|c|c|c|} \hline\hline && i
&\multicolumn{3}{c|}{ii} &\multicolumn{5}{c|}{iii} \\ \hline & $K$
& &-0.003&-0.07&-0.25&0.06&0.37&1.44&5.76&$\infty$\\ \hline $l$
&&&&&&&&&&\\
 2 && 67 & 81 & 63 & 90 & 52 & 46 & 37 &
\textbf{98} & 35
\\ 4 && \textbf{97} & 68 & 41 & 12 & 79 & 89 & 28 & 69 &
\textbf{100} \\ 6 && \textbf{100} & 18 & 66 & 87 & 25 &
4 & \textbf{98} & \textbf{95} & \textbf{100} \\ 8 &&
\textbf{100} & \textbf{98} & 69 & 8 & 74 & 87 &
\textbf{99} & \textbf{100} & \textbf{100} \\ 10 && \textbf{100}
& 34 & 66 & 12 & 44 & 32 & 65 & \textbf{99} &
\textbf{100}
\\ 12 && \textbf{100} & 37 & 80 & 71 & 58 & 7 & 55 & 72 &
\textbf{100} \\ 14 && \textbf{100} & 22 & 29 & 85 & 79 &
40 & 28 & 65 & \textbf{100} \\ 16 && \textbf{100} & 48 &
78 & 88 & 43 & 78 & 89 & \textbf{99} & \textbf{100}
\\ 18 && \textbf{100} & 8 & 5 & 23 & 84 & 19 & \textbf{100} &
\textbf{100} & \textbf{100} \\ 20 && \textbf{100} & 85 & 55 &
\textbf{99} & 53 & 27 & 32 & \textbf{99} & \textbf{100}
\\\hline\hline
\end{tabular}
\end{center}
\caption{ \label{tab:nGpdr}Results for non-Gaussian skies based on
phase differences between consecutive values of $m$ at a fixed
$l$. The results show the significance levels of detected
departures from random phases for simulated distributions
corresponding to (i) coupled phases, (ii) a cardioid distribution
and (iii) a wrapped Cauchy distribution. In the latter two cases
various parameter choices are shown as described in the text.
These results are obtained from 1000 Monte Carlo skies.}
\end{table}

Phases were drawn from wrapped Cauchy distributions in order to
test the method against skies with larger values of kurtosis. The
wrapped Cauchy distribution has a probability density function
given by
\begin{equation}
\label{wrap}
f(\theta)=\frac{1}{2\pi}\;\frac{1-\rho^2}{1+\rho^2-2\rho\cos(\theta)},\quad
0\leq\theta<2\pi,0\leq\rho\leq1
\end{equation}
The kurtosis is given by
\begin{equation}
\label{kwrap} K=\frac{(\rho^2-\rho^4)}{(1-\rho)^2}.
\end{equation}
As \(\rho\rightarrow0\) the distribution tends towards a uniform
distribution. As \(\rho\rightarrow1\) the distribution tends
towards a distribution concentrated at \(\phi=0\). Fisher (1993)
gives a method for simulating the distribution. Sets of
coefficients were created for \(\rho=0,0.2,0.4,0.6,0.8,1\). The
results are given in panel (iii) of Tables \ref{tab:nGpr} and
\ref{tab:nGpdr}. Again \(\rho=0\) is not displayed, but was used
to check the distribution simulation was correct because it should
correspond to a uniformly random distribution. For \(\rho\geq0.6\)
the coefficients are distinctly non--Gaussian. The results in the
Table indicate that the method can pick out departures from
uniformly random phase distributions, at 95\% confidence level for
\(|K|\ga 1\). Note also that the results from \(\rho=1\)
demonstrate that firm conclusions about the Gaussianity and/or
stationarity of the sky are difficult to draw using the quadrupole
alone. Even for the highly non--Gaussian distribution with
$K=\infty$, our statistic does not prove to be discriminatory. In
summary, however, the tests do show that the method can pick out
departures from the random phase hypothesis in CMB sky maps using
low multipole modes.

\subsection{Quadratic Non-Gaussian Maps}

In order to further examine the performance of our statistic, we
applied it to non-Gaussian models with stronger physical
foundations. At large scales, the CMB is dominated by the
Sachs-Wolfe (Sachs \& Wolfe 1967) effect so that the temperature
fluctuations are proportional to fluctuations in the gravitational
potential on the last scattering surface. Non-Gaussian CMB skies
can be generated by introducing a non-linear term in the
gravitational potential $\Phi(n)$
\begin{equation}
\Phi(n)=\Phi_L(n)+f_{nl}(\Phi^2_L(n)-<\Phi^2_L(n)>)
\end{equation}
where $\Phi_L(n)$ refers to the linear part of the potential,
$f_{nl}$ is the non-linear parameter controlling the amount of
non-Gaussianity and the brackets indicate a volume average
(Liguori, Matarrese \& Moscardini 2003). We applied our method to
these types of skies with varying levels of non-Gaussianity
controlled by the parameter $f_{nl}$.

The results for the most extreme form of non-Gaussianity
($f_{nl}=\infty$) are shown in panel (i) in Tables \ref{tab:wmapl}
and \ref{tab:wmapm}. Note that our statistic is clearly unsuitable
for detecting this form of non-Gaussianity. Even for this level of
non-Gaussianity the method is incapable of finding any deviations
of the phases or phase differences from the random phase
hypothesis. The reason for this is the very simple form of phase
association we are testing here does not pick up very sensitively
the quadratic correlations manifested by the model given in
equation (40); see Watts \& Coles (2003). We discussed this in
Section 2.1, in the context of Fourier phases. The reason for this
failure is that the method we are using is much more sensitive to
departures from statistical homogeneity over the celestial sphere
than it is to statistically homogeneous non--Gaussianity. To put
this another way, our method will only be sensitive to departures
from Gaussian behaviour, such as localized edges or blobs, if
these features occupy a substantial fraction of the celestial
sphere. This result will be useful later on, when we interpret the
results we obtain from real data.

It is also worth mentioning that it would not surprising if a
large significance level were attached to one of the $D_l(m)$
values even if the null hypothesis were correct. Since we are
looking at 18 values of $m$ when scanning across fixed values of
$m$, statistically speaking, one value of $m$ should lead to a
probability value larger than 95 per cent. We will not try to
assess the significance level of {\em any} departure from the null
hypothesis in the context of this result as our motivation for
introducing this form of non--Gaussianity is basically to show
that our method struggles to detect it. Whatever method one uses
to combine probabilities across modes would clearly produce a
marginal result in this case for the probability that there is any
departure at all. We will explore this question in more detail in
future work (Dineen et al., in prep.).

\begin{table}
\begin{center}
\begin{tabular}{|cc|c|c|c|c|c|c|c|c|c|} \hline\hline
&& \multicolumn{4}{c|}{$\phi_m(l)$}& &\multicolumn{4}{c|}{$D_m(l)$}\\
$l$&&i&ii&iii&iv&&i&ii&iii&iv\\\hline
2&&89&25&1&7&&42&19&1&12\\
4&&67&56&55&69&&10&43&40&43\\
6&&13&93&94&94&&89&94&94&91\\
8&&88&31&9&31&&31&35&38&30\\
10&&69&12&12&30&&88&16&46&43\\
12&&81&\textbf{97}&89&73&&21&94&81&90\\
14&&53&77&94&94&&93&\textbf{95}&\textbf{96}&\textbf{96}\\
16&&76&\textbf{100}&\textbf{100}&\textbf{99}&&92&\textbf{97}&\textbf{97}&94\\
18&&43&47&84&67&&56&19&86&68\\
20&&23&63&26&5&&5&24&19&7\\
\hline\hline
\end{tabular}
\end{center}
\caption{ \label{tab:wmapl}Results based on phases and
phase differences between consecutive values of $m$ at a fixed
$l$. The results show the significance levels of detected
departures from random phases for (i) the quadratic non-Gaussian map ($f_{nl}=\infty$), (ii) the ILC map, (iii) the cleaned TOH map and (iv) the Wiener filtered TOH map.}
\end{table}

\begin{table}
\begin{center}
\begin{tabular}{|cc|c|c|c|c|} \hline\hline
&& \multicolumn{4}{c|}{$D_l(m)$}\\
$m$&&i&ii&iii&iv\\ \hline
1&&69 &31& 59& 10\\
2&&92 &75& 36& 46\\
3&&63 &23& 35& 67\\
4&&37 &81& 72& 44\\
5&&43 &4& 2& 74\\
6&&10 &4& 46& 34\\
7&&\textbf{97} &79& 87& 55\\
8&&59 &29& 62& 43\\
9&&49 &68& 54& 18\\
10&&84 &17& 79& 81\\
11&&62 &41& 37& 85\\
12&&88 &6& 85& 78\\
13&&25 &76& 65& 60\\
14&&58 &2& 53& 49\\
15&&76 &50& 35& 15\\
16&&86 &48& 16& 28\\
17&&94 &24& 71& 84\\
18&&39 &82& 63& 49\\
\hline\hline
\end{tabular}
\end{center}
\caption{\label{tab:wmapm}Results based on phase differences between consecutive values of $l$ at a fixed
$m$. The results show the significance levels of detected
departures from random phases for (i) the quadratic non-Gaussian map ($f_{nl}=\infty$), (ii) the ILC map,
(iii) the cleaned TOH map and (iv) the Wiener filtered TOH map.}
\end{table}

\subsection{The COBE DMR data}

We now discuss the application of our method  to real data in
order  to see if any signs of non-random phases could be found in
published sky maps. In this subsection we look at the $a_{l,m}$
obtained from COBE-DMR data and in the next subsection we apply
the method to $a_{l,m}$ obtained from three all-sky maps derived
from WMAP data.

The DMR (Differential Microwave Radiometer) instrument on board
the COBE satellite ushered in the era of precision cosmology by
measuring primordial temperature fluctuations of order \(\Delta
T/T=10^{-5}\) in the cosmic microwave background (CMB) radiation.
The DMR instrument comprised six differential microwave
radiometers: two nearly independent channels, labelled A and B, at
frequencies 31.5, 53 and 90 GHz. The data from all three
frequencies have been used to calculate the $a_{l,m}$
corresponding to the CMB signal from regions outside the Galactic
cut. To keep the presentation brief we discuss the behaviour of
$a_{l,m}$ for only the even modes.

The results of the application of the method on the $a_{l,m}$ are
displayed in Table \ref{tab:cobe}. The method was applied to five
realisations (rotations) of the DMR sky in order to examine
whether the results were stable and consequently whether the
conclusions were robust. In the Table it is clear that the large
values of $\chi^2$ (and consequently large significance levels) do
indeed appear stable but lower values (when the phases appear
random) the probabilities depend stronly upon the realisation.
However, we are only interested in significance levels that are
larger than 95\% and these appear stable for all realisations.

\begin{figure} {
\centering{\epsfig{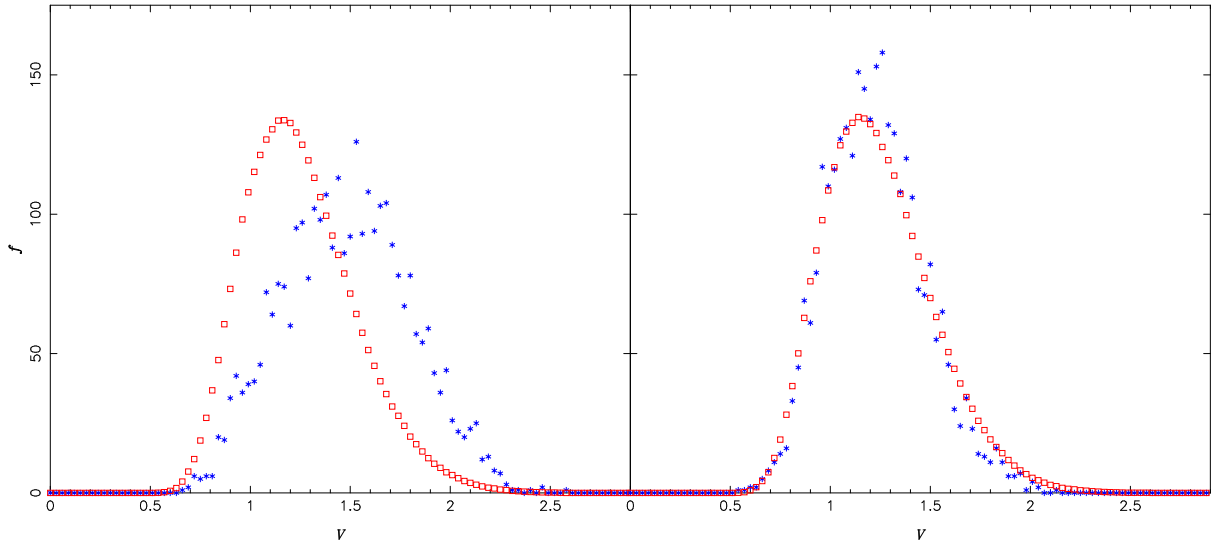}} }
\caption{\label{fig:tentwelve} Examples of the distribution of
Kuiper's statistic \textit{V} for $D_m(10)$ (\textit{left}) and
$D_m(12)$ (\textit{right}) in the COBE-DMR data. The red squares
and blue stars correspond to the average distribution of
\textit{V} for the 1000 MC skies and the distribution for the DMR
data respectively.}
\end{figure}

The statistic indicates that both the phases and phase differences
of $l=10$ mode appear inconsistent with the random phase
hypothesis. For all 5 realisations, the $\chi^2$ values are larger
than 95\% of the MC skies. Figure \ref{fig:tentwelve} shows the
distribution of $V$ obtained for realisation 1 from Table
\ref{tab:cobe} compared with the average distribution of the MC
skies for $l=10$ and 12. The distribution of $D_m(10)$ clearly
differs from that of the random phase MC skies whereas $D_m(12)$
appears by eye to be drawn form the same distribution.
 The results in Table \ref{tab:cobe} also show that the phases obtained from the $l=18$ also appear non-random,
  but the confidence of this result is dependent on the realisation.

\begin{table}
\begin{center}
\begin{tabular}{|ccccccccccccc|} \hline\hline
&& \multicolumn{5}{c|}{$\phi_m(l)$}& &\multicolumn{5}{c|}{$D_m(l)$}\\
$l$&&1&2&3&4&5&&1&2&3&4&5\\ \hline
2&&35&40&46&46&47&&38&38&45&40&34\\
4&&81&83&81&83&79&&78&76&76&78&78\\
6&&49&31&52&53&76&&46&44&43&37&41\\
8&&19&43&14&41&1&&68&63&67&65&63\\
10&&\textbf{97}&\textbf{97}&\textbf{97}&\textbf{98}&\textbf{97}&&\textbf{96}&\textbf{96}&\textbf{96}&\textbf{96}&\textbf{96}\\
12&&49&38&1&1&59&&13&24&14&3&7\\
14&&61&77&33&31&35&&3&10&77&8&10\\
16&&4&77&87&92&39&&57&66&67&59&65\\
18&&\textbf{97}&94&94&\textbf{96}&\textbf{97}&&92&92&91&94&93\\
20&&64&58&7&51&36&&30&19&18&30&23\\\hline\hline
\end{tabular}
\end{center}
\caption{\label{tab:cobe}Results for COBE DMR data based on phases and
phase differences between consecutive values of $m$ at a fixed
$l$. The method was applied to 5 realisations of the DMR sky in order to demonstrate the stability of the results.}
\end{table}

 The issue of non-Gaussianity in the COBE-DMR data has been
 controversial (Ferreira, Magueijo \& Gorski 1998; Bromley \& Tegmark 1999; Magueijo
 2000), but is likely to involve systematic problems identified in
 the data set we use. Note however that the strong detection we
 have found is at a different harmonic mode $l=10$ than that
 originally claimed by Ferreira et al. (1998) who used a
 diagnostic related to the bispectrum. There need not be a
 conflict here because, as we discussed above, we are not
 sensitive  to the same form of phase correlations in this method as the
 bispectrum.

 We do not wish to go further into the statistical properties of
 the COBE-DMR data here. The results we have presented are from
 the original data which is now known to have suffered from some systematic problems
 (Banday, Zaroubi \& Gorski 2000). The effect we are seeing is therefore probably not
 primordial, making this largely an academic discussion. The data have also partly been superseded by WMAP.
 It does, however, illustrate the importance of using complementary approaches
 to identify all possible forms of behaviour in data sets even
 when they are of experimental, rather than cosmological, origin.

\subsection{WMAP results}
The WMAP instrument comprises 10 differencing assemblies
(consisting of two radiometers each) measuring over 5 frequencies
(\(\sim \)23, 33, 41, 61 and 94 GHz). The WMAP team have released
an internal linear combination (ILC) map that combined the five
frequency band maps in such a way to maintain unit response to the
signal CMB while minimising the foreground contamination. The
construction of this map is described in detail in Bennett et al.
(2003). To further improve the result, the inner Galactic plane is
divided into 11 separate regions and weights determined
separately. This takes account of the spatial variations in the
foreground properties. The final map covers the full-sky and
should represent only the CMB signal, although it is not
anticipated that foreground subtraction is perfect.

The WMAP ILC map is issued with estimated errors in the map-making
technique. This uncertainty could be included in the Monte Carlo
simulations of the null hypothesis used to construct sampling
distributions of our test statistic. This would be difficult to do
rigorously as errors are highly correlated, but is possible in
principle. We have decided not to include the experimental errors
in this analysis because it is just meant to illustrate the
method: a more exhaustive search for CMB phase correlations (for
which such considerations would be relevant) will have to wait for
cleaner maps than are available now.

Following the release of the WMAP 1 yr data Tegmark, de
Oliveira-Costa \& Hamilton (2003; TOH) have produced a cleaned CMB
map. They argued that their version contained less contamination
outside the Galactic plane compared with the internal linear
combination map produced by the WMAP team. The five band maps are
combined with weights depending both on angular scale and on the
distance from the Galactic plane. The angular scale dependence
allows for the way foregrounds are important on large scales
whereas detector noise becomes important on smaller scales. TOH
also produced a Wiener filtered map of the CMB that minimisizes
rms errors in the CMB. Features with a high signal-to-noise are
left unaffected, whereas statistically less significant parts are
suppressed. While their cleaned map contains residual Galactic
fluctations on very small angular scales probed only by the W
band, these fluctuations vanish in the filtered map.

The three all-sky maps are available in
HEALPix\footnote{http://www.eso.org/science/healpix/} format
(Gorski, Hivon \& Wandelt 1999). We derived the $a_{l,m}$ for each
map using the {\tt anafast} routine in the HEALPix package. The
results are shown in panels (ii) to (iv) in Tables \ref{tab:wmapl}
and \ref{tab:wmapm}. The results for the ILC map suggest there are
departures from the random phase hypothesis for the phases of
$l=12$ and 16 and the phase differences for $l$=14 and 16. The
departure for $l$=16 appears very strong with the $\chi^2$ value
being larger than 995 of the values obtained for the MC skies. For
these data we have also explored the effect of measuring
differences between the phases at different $l$ for the same value
of $m$. (We did not present results in this case for the COBE-DMR
data, as we only discussed the even $l$-modes in that case and
differences between adjacent $l$ would require every mode.) Notice
that the confidence levels of non-uniformity are larger for $D_m$
at fixed $l$ than for $D_l$ at fixed $m$. This is an interesting
feature of the WMAP data.

\begin{figure} {
\centering{\epsfig{file=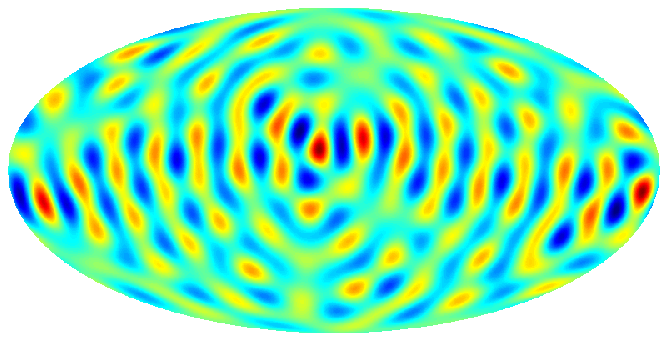,width=14cm}}
\centering{\epsfig{file=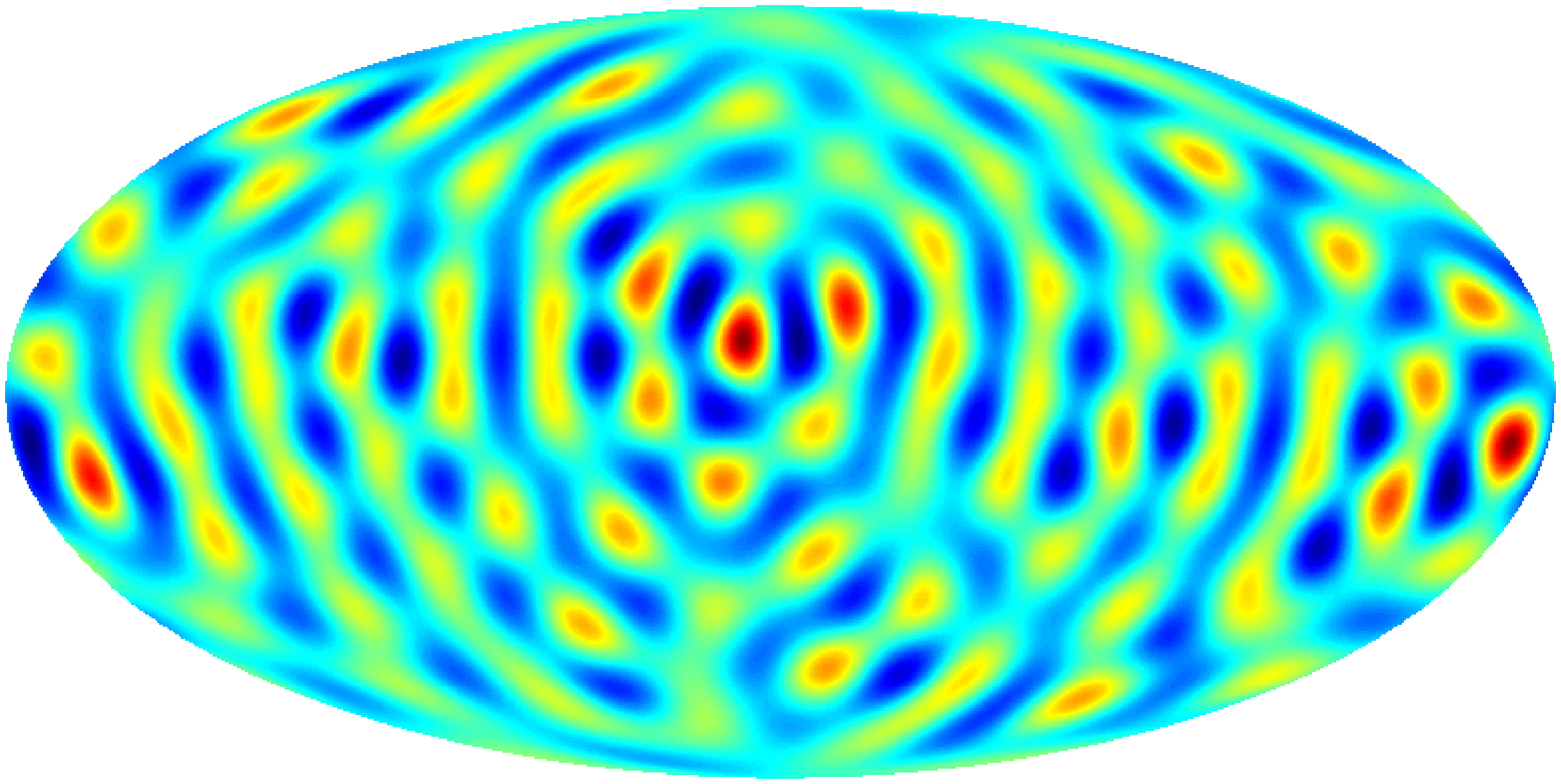, width=14cm}}
\centering{\epsfig{file=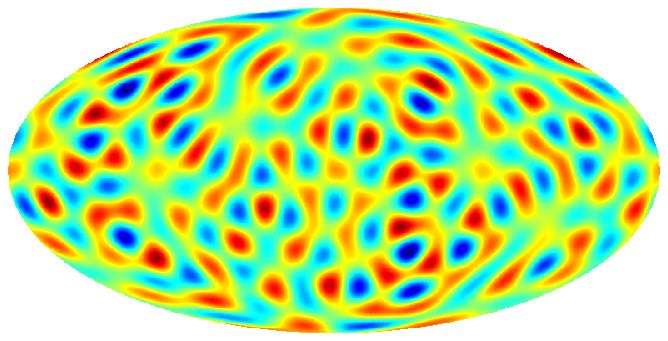,width=14cm}}}
 \caption{\label{fig:ilc16}\textit{Top}: Internal linear
combination map constructed with $a_{l,m}$ for $l=16$ only.
\textit{Middle}: The TOH Wiener-filtered map  with $a_{l,m}$ for
$l=16$ only. \textit{Bottom}: A random-phase realization of the
map including modes with $l=16$ only.}
\end{figure}

To investigate this mode further we therefore reconstructed the
temperature pattern on the sky solely using only the spherical
harmonic modes for $l=16$ with their appropriate $a_{l,m}$  and
compared this with a map with the same power spectrum but with
random phases (figure \ref{fig:ilc16}). The result is a striking
confirmation of the above argument that our method is very
sensitive to departures from statistical homogeneity. The $l=16$
modes are clearly forming a more structured pattern than the
random-phase counterpart. The alignment of these modes with the
Galactic plane is striking; it may be that the angular scale
corresponding to $l=16$ may relate to the width of the zones used
to model the Galactic plane.

The two maps produced by Tegmark et al. (2003; TOH) also suggest
that the phases for the $l=16$ scale are non-random. The cleaned
map shows departures from the random phase hypothesis at greater
than 95\% confidence for the phases of $l$=16 and phase
differences of $l=14$ and 16. The Wiener filtered map finds
departures for the phases corresponding to $l=16$ and phase
differences in $l$=14. Both the cleaned and the Wiener filtered
maps of TOH display similar band patterns to that observed in the
$l=16$ map for the ILC; we show the Wiener filtered TOH map for
comparison.

The morphological appearance of these fluctuations as a band
across the Galactic plane may be indicative of residual foreground
still present in the CMB map after subtraction. This is the most
likely explanation for the alignment of features relative to the
Galactic coordinate system, since the scanning noise is not
aligned in this way. Further evidence for this is provided by
Naselsky et al. 2003 from cross--correlations of the phases of CMB
maps with various foreground components. In any case one should
certainly caution against jumping to the conclusion that this
represents anything primordial. These are preliminary data with
complex variations in signal-to-noise across the sky.
Nevertheless, the results we present here do show that the method
we have presented is useful for testing realistic observational
data for departures from the standard cosmological statistics.

\section{Discussion and Conclusions}

In this paper we have presented a new method of testing CMB data
for departures from the homogeneous statistical behaviour
associated with Gaussian random fields generated by inflation. Our
method uses some of the information contained within the phases of
the spherical harmonic modes of the temperature pattern and is
designed to be most sensitive to departures from stationarity on
the celestial sphere. The method is relatively simple to
implement, and has the additional advantage of being performed
entirely in ``data space''. Confidence levels for departures from
the null hypothesis are calculated forwards using Monte Carlo
simulations rather than by inverting large covariance matrices.
The method is non-parametric and, as such, makes no particular
assumptions about data. The main statistical component of our
technique is a construction known as Kuiper's statistic, which is
a kind of analogue to the Kolmogorov-Smirnov test, but for
circular variates.

In order to illustrate the strengths and weaknesses of our
approach we have applied it to several test data sets. To keep
computational costs down, but also to demonstrate the usefulness
of non-parametric approaches for small data sets, we have
concentrated on low--order spherical harmonics.

We first checked the ability of our method to diagnose non-uniform
phases of the type explored by Watts, Coles \& Melott (2003) and
Matsubara (2003). The method is successful even for the low--order
modes that have few independent phases available. We then applied
the method to  quadratic non--Gaussianity fields of the form
discussed by Watts \& Coles (2003) and Liguori, Matarrese \&
Moscardini (2003). Our method is not sensitive to the particular
form of phase correlation displayed by such fields as, although
non--Gaussian, they are statistically stationary. Phase
correlations are present in such fields, but do not manifest
themselves in the simple phase or phase-difference distributions
discussed in this paper. These two examples demonstrate that our
method is a useful diagnostic of statistical non-uniformity and
its applicability to non--Gaussian fields is likely to be
restricted to cases where the pattern contains strongly localized
features, such as cosmic strings or textures (e.g. Contaldi et al.
2000).

We next turned our attention to the COBE--DMR data. Claims and
counter-claims have already been made about the possible
non--Gaussian nature of these data, the most likely explanation of
the observed features being some form of systematic error. Our
test does show up non-uniformity in the phase distribution at high
confidence levels for the $l=10$ mode and, less robustly, at
$l=18$. This is a different signal to that claimed previously to
be indicative of non--Gaussianity. Its interpretation remains
unclear.

Finally we applied our method to various representations of the
WMAP preliminary data release. In this case we find clear
indications at $l=16$ and, less significantly, at $l=14$. The
distribution of the $l=16$ harmonics on the sky shows that this
result is not a statistical fluke. The data are clearly different
from a random--phase realisation with the same amplitudes and
there is certainly the appearance of some correlation of this
pattern with the Galactic coordinate system. We are not claiming
that this proves there is some form of primordial non--Gaussianity
in this dataset. Indeed the WMAP data come with a clear warning
its noise properties are complex so it would be surprising if
there were no indications of this in the preliminary data. It will
be useful to apply our method to future releases of the data in
order to see if the non-uniformity of the phase distribution
persists as the signal-to-noise improves. At the moment, however,
the most plausible interpretation of our result is that it
represents some kind of Galactic contamination, consistent with
other (independent) claims (Dineen \& Coles 2003; Chiang et al.
2003; Eriksen et al. 2003; Naselsky et al. 2003).

As a final comment we mention that the aptitude of our method for
detecting spatially localised features (or departures from
statistical homogeneity generally) suggests that it is may be
useful as a diagnostic of the repeating fluctuation pattern
produced on the CMB sky in cosmological models with compact
topologies (Levin, Scannapieco \& Silk 1998; Scannapieco, Levin \&
Silk 1999; Rocha et al. 2002); see Levin (2002) for a review. We
shall return to this issue in future work.

\section*{Acknowledgements.}

We acknowledge the use of the Legacy Archive for Microwave
Background Data Analysis (LAMBDA). Support for LAMBDA is provided
by the NASA Office of Space Science. The COBE datasets were
developed by the NASA Goddard Space Flight Centre under the
guidance of the COBE Science Working Group and were provided by
the NSSDC. We acknowledge NASA and the WMAP Science Team for their
data products. We thank Joao Magueijo for supplying us with the
spherical harmonic coefficients he used for the COBE 4-year data
(Magueijo 2000). We also thank Michele Liguori for letting us use
his non--Gaussian simulations. We are grateful to Pedro Ferreira,
Sabino Matarrese and Mike Hobson for useful discussions related to
this work. A preliminary version of some of the work presented in
this paper was performed by John Earl and Dean Wright as an
undergraduate project, which subsequently won a prize awarded by
the software company Tessella. Patrick Dineen is supported by a
University Research Studentship. This work was also partly
supported by PPARC grant number PPA/G/S/1999/00660.

\end{document}